\documentstyle[osa]{revtex}

\begin{document}

\newcommand{\beq}{\begin{equation}}  
\newcommand{\eeq}{\end{equation}}
\newcommand{\bqa}{\begin{eqnarray}} 
\newcommand{\eqa}{\end{eqnarray}}
\newcommand{\nn}{\nonumber} 
\newcommand{\erf}[1]{Eq.~(\ref{#1})}
\newcommand{\dg}{^\dagger}
\newcommand{\smallfrac}[2]{\mbox{$\frac{#1}{#2}$}}
\newcommand{\bra}[1]{\langle{#1}|} 
\newcommand{\ket}[1]{|{#1}\rangle}
\newcommand{\ip}[1]{\langle{#1}\rangle}
\newcommand{\sch}{Schr\"odinger } 
\newcommand{\schs}{Schr\"odinger's }
\newcommand{\hei}{Heisenberg } 
\newcommand{\heis}{Heisenberg's }
\newcommand{\half}{\smallfrac{1}{2}} 
\newcommand{\bl}{{\bigl(}}
\newcommand{\br}{{\bigr)}} 
\newcommand{\ito}{It\^o }
\newcommand{\str}{Stratonovich } 

\title{\Large {\bf Extending Heisenberg's measurement--disturbance relation to the 
twin-slit case.}}
\author{\large H.\ M.\ Wiseman\footnote{\normalsize for correspondence: tel: +61 
7 3365 3425. fax: +61 7 3365 1242. e-mail: wiseman@physics.uq.edu.au} }
\address{\large Department of Physics, The University of Queensland, St.\ Lucia 4072, 
Australia }

\date{Foundations of Physics {\bf 28}, 1619 (1998)}
 
\maketitle

{\large 

\begin{abstract}
	{\large Heisenberg's position-measurement--momentum-disturbance relation is 
	derivable from the 
	uncertainty relation $\sigma(q)\sigma(p) \geq \hbar/2$ 
	only for the case when the 
	particle is initially in a momentum eigenstate. Here I derive a new 
	measurement--disturbance relation which applies when the particle is 
	prepared in a twin-slit superposition and the measurement can
	determine at which slit the particle is present. The relation is 
	$d \times \Delta p \geq 2\hbar/\pi$, where $d$ 
	is the slit separation 
	and $\Delta p=D_{M}(P_{f},P_{i})$ is the Monge distance 
	between the initial $P_{i}(p)$ and final $P_{f}(p)$
	momentum distributions.}
\end{abstract}



\section{\large {\bf Introduction}}

There is a fundamental ambiguity in Heisenberg's uncertainty 
relation
which dates back to its birth in the famous 1927 paper 
\cite{Hei27}. Here Heisenberg introduced the relation 
in the context of a position measurement by a $\gamma$-ray 
microscope, as (in my notation)
\beq \label{mdr}
\epsilon_{q} \times \delta p  \sim h,
\eeq
where $\epsilon_{q}$ is ``the precision with which the value $q$ is known (say 
the mean error of $q$)'' and $\delta p $ is ``the discontinuous 
change of $p$ in the Compton effect.'' By 
``mean error'' Heisenberg evidently meant root-mean-square error, or its 
equivalent, and I will follow this use. The relation (\ref{mdr}) 
we may call the 
Heisenberg measurement--disturbance relation. As Heisenberg says, 
\begin{quote}
	The instant the position is determined \ldots the electron undergoes 
	a discontinuous change in momentum. This change 
	is the greater the \ldots more exact the determination of 
	the position.
\end{quote}
The roles of $p$ and $q$ in this description are clearly not symmetric.
But in the same work Heisenberg talks about the uncertainty 
relation as referring to ``simultaneous determination of two 
canonically conjugate quantities'', which is a different statement. 
Not long after Heisenberg, Weyl \cite{Wey28} 
put this latter statement on a rigorous footing as
\beq \label{ur}
\sigma(q)\sigma(p) \geq \hbar /2.
\eeq
 Here $\sigma(q),\sigma(p)$ are the 
simultaneous values of the standard deviations of $q,p$. Following 
modern use, I will call 
this an uncertainty relation.

What link is there between the uncertainty relation (\ref{ur}) and 
Heisenberg's measurement--disturbance relation (\ref{mdr})? 
It seems fairest to
 let Heisenberg speak for himself. 
In his most complete description of his position, contained in his 
1930 book \cite{Hei30}, he first gives a derivation of the relation
(\ref{ur}). This derivation, using only ``the 
mathematical scheme of quantum theory and its physical interpretation''
says nothing about
momentum transfer or position measurement. It is only in the next 
section, ``Illustrations of the Uncertainty Relations'', that he 
introduces these ideas, and does so very carefully:
\begin{quotation}
	The uncertainty principle refers to the degree of indeterminacy in 
	the possible present knowledge of the simultaneous values of various 
	quantities with which quantum theory deals; it does not restrict, for 
	example, the exactness of a position measurement alone or a velocity 
	measurement alone. Thus suppose that the velocity of a free electron 
	is known, while the position is completely unknown. Then the principle 
	states that any subsequent observation of the position will alter the 
	momentum by an unknown and indeterminable amount such that  
	after carrying out the experiment our knowledge of the 
	electronic motion is restricted by the uncertainty relation.
\end{quotation}

As Heisenberg appeared to be well aware, this
is the only statement about momentum disturbance and position 
measurement error which one can make as 
a logical consequence of the uncertainty relation (\ref{ur}). The 
standard deviations $\sigma(p)$ and $\sigma(q)$ refer to the state of 
the particle after the measurement. It is 
only because the particle state prior to the measurement was
a momentum eigenstate (having zero dispersion in momentum) that one can 
equate $\sigma(p)$ with $\delta p $, the mean momentum 
disturbance. Likewise it is only because the particle prior to the 
measurement had a completely undefined position that one can equate
$\sigma(q)$ with the mean error $\epsilon_{q}$ of the position 
measurement. To see this latter point, consider the case where the particle is 
not in a momentum eigenstate, but instead localized at two narrow 
slits of width $a$, separated by a distance $d$. Then a position 
measurement with an error of order $d$ will resolve the two slits, and the 
particle will become localized at one of them. The resulting final standard 
deviation in the position $\sigma(q) \sim a$ is not related to the 
error of the measurement $\sim d$.

In the derivation of \erf{ur} in Ref.~\cite{Hei30}, Heisenberg 
does not explicitly use the commutation relations
\beq
[q,p] = i\hbar,
\eeq
but rather the Fourier-transform relation between the position and 
momentum representations. It should especially be noted that there is 
no hint in Heisenberg's work 
that the more general Robertson uncertainty relation \cite{Rob29}
\beq \label{Rur}
\sigma(A)\sigma(B) \geq \frac{1}{2}\left|\ip{[A,B]}\right|
\eeq
leads to a more general measurement--disturbance relation. Instead, there 
are good reasons for maintaining that it does not. If, {\em unlike} $p$ 
and $q$, the quantities 
$A$ and $B$ are {\em not} canonically conjugate, then 
preparing the system in an eigenstate of $A$ need not
ensure that all 
values of $B$ are equally likely prior to the measurement.  After the 
measurement, the standard deviation $\sigma(A)$ 
can still be identified with the disturbance in the quantity $A$ 
caused by the measurement. However the 
quantity $\sigma(B)$ is not 
necessarily the accuracy of the measurement of $B$, because there may 
have been some information regarding the  
value of $B$ prior to the measurement
(the argument follows the same 
lines as given above for the twin-slit case). The lesson is 
that the Heisenberg measurement--disturbance relation is of quite 
particular content.

In his actual illustrations in the 1930 book (including the famous 
$\gamma$-ray microscope from the 1927 paper), Heisenberg 
seeks to give an intuitive interpretation of the momentum transfer 
(in terms of the Compton recoil and such like). But he never claims 
to rigorously prove any momentum--disturbance relation other than 
\beq \label{hur}
\epsilon_{q} \times \delta p  \geq \hbar/2
\eeq
which applies when the particle 
is initially in a momentum eigenstate (or at least a state with 
negligible momentum dispersion).

The problem is that one often wishes to consider position 
measurements, and hence momentum transfers, in situations in which 
the particle is not in a momentum eigenstate.  A particular example 
of interest which I have already mentioned 
is that of the twin slits. This was one of the subjects of the 
Bohr--Einstein debates \cite{BohrEinst} and, more recently, has been 
surrounded by the controversy over  
whether there is a momentum transfer (of order commensurate with 
the uncertainty relation) concomitant with determining which slit a 
particle passes through. 
Scully, Englert and Walther \cite{ScuEngWal91}  
prove that if the particle is already localized at one of the slits 
then there need be no such momentum transfer. On this basis they 
say that 
it is particle-wave complementarity, rather than the uncertainty principle, 
which explains the loss of the interference pattern when one 
measures which slit the particle went through. Storey, Tan, Collett and 
Walls \cite{StoTanColWal94} on the contrary have 
claimed that there is always a transverse 
(that is, in the direction of the line connecting the slits)
momentum disturbance at least equal to $\hbar/d$, where
$d$ is the slit separation. In this they uphold the opinion of Bohr 
\cite{BohrEinst} 
that one can regard complementarity as being enforced by the 
uncertainty principle. Further exchanges are found in 
Refs.\cite{QI94,Nature95}.

It was pointed out by myself and Harrison 
\cite{WisHar95} that the basis of the disagreement lay in 
a difference over the definition of momentum transfer. As noted above,
the momentum transfer is in general 
defined unambiguously only if the particle is initially in a 
momentum eigenstate. The calculations of Scully {\em et al.}
concern the {\em local} momentum transfer. This is the momentum 
transfer which can be seen in the shift or 
broadening of the momentum distribution 
of a particle already localized at one of the slits. By contrast, 
the momentum transfer distribution considered by Storey {\em et al.} 
can perhaps be best characterized as the {\em potential} momentum 
transfer. It would be an actual momentum transfer if the particle were 
 initially in a momentum eigenstate.

In Ref.~\cite{Wis97a} myself and Harrison, 
together with Collett, Tan, Walls 
and Killip, showed that by using the Wigner function formalism 
one can identify different types of momentum transfer, which we 
called local and nonlocal. The local momentum transfer corresponds to 
the concept of momentum disturbance used by Scully, Englert and 
Walther \cite{ScuEngWal91}. We showed, in agreement with the 
claims of Scully {\em et al.}, that this may indeed be zero and 
that the momentum transfer in the theorem of Storey {\em et 
al.} was not relevant to this calcution. 
On the other hand, we showed that a 
particular measure of the nonlocal momentum 
transfer is always greater than $\pi\hbar/2d$, and this
is derived in the same
manner as the theorem of Storey {\em et al.} \cite{StoTanColWal94}. 
It is this nonlocal momentum transfer which caused (in the Wigner function 
formalism) the loss the loss of the interference fringes.

In this work I want to revisit this question afresh. Rather than 
concentrating on the issue of interference and the loss of it, I 
propose to look at the question in the following context. As noted 
above, Heisenberg 
was the first to derive a rigorous 
measurement--disturbance relation (\ref{hur}), in the case 
when a particle is initially in a momentum 
eigenstate.
Now consider a different situation. The particle, rather than having 
an equal probability amplitude of being at all points $q$, now has an 
equal probability of being at just two points (or at least two 
identical small regions) separated by a distance $d$. A measurement 
which can distinguish between these two regions (hereafter known as 
slits) must have a discrimination length scale $\sim d$. 
 Therefore we expect that if any extension of Heisenberg's 
measurement--disturbance relation is possible, the particle's momentum 
should be disturbed by an amount $\sim \hbar/d$. 

The aim of 
this paper is to show that it is indeed possible to derive 
a relation of this sort.  
I would not call the relation I derive a Heisenberg 
relation, because Heisenberg's measurement--disturbance relation is 
based on the so-called Heisenberg 
uncertainty relation (\ref{hur}). The extension of this 
relation to the twin-slit case need not be based on that particular 
theorem, in particular because the measure of the momentum transfer 
cannot be based on the standard deviation. 
Nevertheless, it will be based in the formalism of quantum 
mechanics, including the conjugate relation of position and 
momentum, just as much as Heisenberg's measurement--disturbance 
relation was. The momentum transfer I calculate in this paper is 
different from any of those in 
Refs.~\cite{ScuEngWal91,StoTanColWal94,Wis97a}. 
Moreover, the measure of momentum transfer I 
propose uses only the momentum distributions of the particle 
before and after the measurement. As such it 
could not be criticized as being merely a potential momentum transfer.
Despite the importance of nonlocal momentum transfer in 
the recent work of Ref.~\cite{BerEng98}, 
the issue of locality or nonlocality 
is irrelevant to this work and will not be discussed.

\section{\large {\bf Describing the Measurement}}

The starting point for the calculation is the initial wavefunction
for the twin-slit case, which can be written as
\beq \label{state}
\psi_{i}(q) = 2^{-1/2}\left[ \phi_{a}(q) + 
\phi_{a}(q-d) \right].
\eeq
Here $\phi_{a}(q)$ is a wavefunction parameterized by 
a positive real number 
$a$ such that its width scales as $a$ and
\beq \label{deltafn}
\lim_{a \to 0} |\phi_{a}(q)|^{2} = \delta(q).
\eeq
For example, 
\beq \label{particf}
\phi_{a}(q) = (2\pi a^{2})^{-1/4}\exp(-q^{2}/4 a^{2})
\eeq
would do, and I will use this form for some specific calculations.
For this example the state (\ref{state}) is normalized only in the limit
$a \to 0$, but that is all that we need.

In the momentum representation (indicated by a tilde), the initial state 
is, up to  an irrelevant phase factor,
\beq \label{imwf}
\tilde{\psi}_{i}(p) =  \tilde{\phi}_{a}(p) \sqrt{2}\,
\cos\frac{pd}{2\hbar},
\eeq
where 
\beq
\tilde{\phi}_{a}(p) = \frac{1}{\sqrt{2\pi\hbar}}\int dq \, 
e^{-ipq/\hbar} \phi_{a}(q) .
\eeq
The momentum probability distribution is therefore
\beq \label{imd}
P_{i}(p) = \left(1 + \cos\frac{pd}{\hbar}\right) 
{\cal E}(p),
\eeq
where 
\beq \label{defcalEp}
{\cal E}(p) = |\tilde{\phi}_{a}(p)|^{2}.
\eeq
It is the oscillations of period $2\pi \hbar/d$ under the envelope 
${\cal E}(p)$ which are evidence of the coherent superposition of the 
particle being at the two slits at $q = 0,d$. 

The effect of a position measurement on the particle's wavefunction is 
to change it into
\beq
\psi_{f}(q) = N_{\xi}^{-1/2}O_{\xi}(q)\psi_{i}(q),
\eeq
where $O_{\xi}(q)$ is a function relating to a particular measurement 
result $\xi$ and
\beq
N_{\xi} = \int dq |O_{\xi}(q)\psi_{i}(q)|^{2}
\eeq
is the probability for obtaining that result \cite{StoTanColWal94}. 
Obviously the sum over 
all probabilities $N_{\xi}$ must equal unity. We are interested in the 
case where the particular result 
$\xi$ successfully distinguishes between the two slits.
Then $O_{\xi}(q)$ must be 
zero in the region of one of the slits. Without loss of generality we 
may take it to be zero for $q \approx d$. Thus the final 
wavefunction is
\beq
\psi_{f}(q) = N_{\xi}^{-1/2}O_{\xi}(q) 2^{-1/2}\phi_{a}(q).
\eeq
Creating a MacLaurin expansion of $\ln O_{\xi}(q)$ yields
\beq \label{psifq}
\psi_{f}(q) \propto \exp(\alpha q + \beta q^{2} + \ldots)\phi_{a}(q),
\eeq
where $\alpha,\beta,\ldots$ are complex numbers.

Taking the particular form of $\phi_{a}(q)$ in \erf{particf},
and ignoring the unwritten higher order terms in \erf{psifq} we get
\beq
\tilde{\psi}_{f}(p)
\propto \exp\left[  (\alpha - 
ip/\hbar)^{2}/(a^{-2}+4\beta) \right].
\eeq
Thus in the limit $a \to 0$ we find the final momentum 
distribution to be
\beq
P_{f}(p) = \lim_{a\to 0} |\tilde{\psi}_{f}(p)|^{2}=
\frac{a}{\hbar}\sqrt\frac{2}{\pi}
\exp\left[-a^{2} (p/\hbar-k)^{2}\right],
\eeq
which has been normalized. Here $k = {\rm 
Im}(\alpha)$. The real part of $\alpha$, and the whole of $\beta$ 
(and also the higher order terms) become irrelevant 
when the limit $a \to 0$ is taken. In fact, the Gaussian form 
of the original wavefunction $\phi_{a}(q)$ is not required for 
this result and we can take the final momentum probability 
distribution to be 
more generally
\beq
P_{f}(p) = {\cal E}(p-\hbar k),
\eeq
where ${\cal E}(p)$ is as defined in \erf{defcalEp}.

\section{\large {\bf Quantifying the Momentum Transfer}}

From the preceding section we see that the momentum probability 
distributions before and after the position measurement which 
distinguishes between the two slits are respectively
\bqa
P_{i}(p) &=& \left(1 + \cos\frac{pd}{\hbar}\right) 
{\cal E}(p), \\
P_{f}(p) &=& {\cal E}(p-\hbar k),
\eqa
where ${\cal E}(p)$ is an envelope which is arbitrarily smooth and 
broad. Obviously the two momentum distributions are different and so 
one would be justified in saying that there must have been some 
momentum transfer. One approach to quantifying this transfer 
would be to compare the 
moments of the two distributions. However, for the case $k=0$, 
the mean and standard deviation of 
$P_{i}(p)$ and $P_{f}(p)$ are identical \cite{Wis97a}. 
This case corresponds to the scheme proposed by 
Scully {\em et al} \cite{ScuEngWal91}, in which the {\em 
local} momentum transfer (which is the only momentum transfer with 
which they 
are concerned) is zero. While comparing the moments of 
$P_{i}(p)$ and $P_{f}(p)$ may identify the local 
momentum transfer (or lack of it) \cite{Wis97a}, it  
is evidently not a good measure of the overall change in the momentum 
distribution.

The question we thus face is, 
what is a better way to quantify momentum transfer? One could seek a 
measure based on the measurement function $O_{\xi}(q)$, as done 
by Storey {\em et al.} \cite{StoTanColWal94}. The problem with this 
approach is that it is independent of the initial state. 
This would seem to imply that there is necessarily a momentum transfer 
even if the particle is already localized at one slit. While not 
logically impossible\cite{WisHar95}, this is hard to 
accept on physical grounds given that Scully {\em et al.} 
\cite{ScuEngWal91,QI94,Nature95} have shown that
the momentum distributions may be unchanged by the 
measurement in that case. This difficulty can be overcome 
by turning to the Wigner function formalism \cite{Wis97a}, as mentioned 
above. Nevertheless the simpler solution would seem to be to abandon 
$O_{\xi}(q)$ altogether and 
seek a (non-moment-based) 
measure which involves only the initial and final momentum 
distributions.

Our question is then very closely related to one considered by the 
18th century French mathematician, Monge \cite{Mon1781}. 
The problem involves the transportation of a mass of 
soil from a given configuration 
(e.g. a heap) to another configuration (e.g. a dike) by haulage. 
 As an idealization, we can assume that 
the mass of soil is divided into an arbitrarily large number of 
identical small loads each of which is 
hauled separately. A particular strategy for shifting the soil will 
therefore be characterized by an average distance over which the 
loads are hauled (vertical displacement being assumed 
negligible). Monge's problem is to minimize this average distance. The 
haulage strategy which achieves this is known as the Monge plan and 
the resulting distance is known as the Monge distance. 
The Monge distance so defined is  a good metric (in the 
mathematical sense) over the 
space of all possible soil distributions. The distance resulting from 
a non-optimal strategy is not a good metric, since inefficient 
workers could move loads of soil backwards and forwards over a large 
distance without changing the overall distribution of soil at all. 

In modern probability theory, the Monge distance is used to define a 
metric over the space of probability distributions \cite{Rac91}.
For simplicity, consider only distributions in one real variable. Imagine 
dividing up the area under the two curves into infinitesimal elements 
of equal area. 
Then the Monge distance is the 
minimum mean distance over which elements of the first 
probability distribution can be shifted so as to transform 
it into the second 
probability distribution. Clearly the Monge distance 
has the same dimension as the variable whose distribution we are 
considering. 
There are generalizations of Monge's distance (such as the 
Fr\'echet distance, which is the minimum root-mean-square distance), 
but there is no particular reason to 
prefer them for this problem.

For the case at hand, the Monge distance between the 
initial $P_{i}(p)$ and final $P_{f}(p)$ 
momentum distributions has the dimensions of momentum, and can 
in fact be identified with the average of the absolute value of the 
momentum transfer by the 
measurement. In this one-dimensional case, 
the Monge plan is to transport the infinitesimal elements along the 
line without changing their order \cite{Rac91}. 
Translating these words into mathematics, 
 the Monge distance $D_{M}$ is given by
\beq \label{DMP}
D_{M}(P_{f},P_{i}) = \int_{0}^{1} d\lambda 
|F^{-1}_{i}(\lambda) - F^{-1}_{f}(\lambda)|.
\eeq
Here $F^{-1}(\lambda)$ is defined by 
\beq
F^{-1}\bl F(p) \br \equiv p,
\eeq
where 
$F(p)$ is the fiducial distribution
\beq
F_{i/f}(p) = \int_{-\infty}^{p} dp' P_{i/f}(p').
\eeq 
By a change of variable \erf{DMP} becomes
\beq \label{defMD}
D_{M}(P_{f},P_{i}) = \int_{-\infty}^{\infty} dp' |F_{i}(p') - F_{f}(p')|.
\eeq

In our case we have
\bqa
F_{i}(p) &=& {\cal F}(p) + \frac{\hbar}{d}{\cal 
E}(p)\sin\frac{pd}{\hbar}  - \frac{\hbar}{d}\int_{-\infty}^{p}  
\sin\frac{p'd}{\hbar}\, d {\cal E}(p') \label{3t}\\
F_{f}(p) &=& {\cal F}(p-\hbar k)
\eqa
 where ${\cal F}(p)$ is the fiducial distribution of ${\cal E}(p)$.
Now since ${\cal E}(p)$ has a width of order $\hbar/a \gg \hbar/d$, 
the size 
of the three terms in \erf{3t} are of order $1, a/d, (a/d)^{2}$ 
respectively. 
As I will show, the first-order term yields a finite
contribution to $D_{M}$, so the second-order term can be neglected.

Substituting the expressions for the fiducial distribution into 
\erf{defMD} yields
\beq
\int_{-\infty}^{\infty} dp' \left| \int_{p'-\hbar k}^{p'}{\cal E}(p'') 
dp'' + \frac{\hbar}{d}{\cal E}(p')\sin\frac{p'd}{\hbar} \right| .
\eeq
Using the fact that $k a,a/d \ll 1$, in the limit $a\to 0$ we can replace this 
expression by
\beq
D_{M}(P_{f},P_{i}) = \int_{-\infty}^{\infty} dp' {\cal E}(p')\left| \hbar k +
\frac{\hbar}{d}\sin\frac{pd}{\hbar} \right| = \left\langle 
\left|\hbar k +
\frac{\hbar}{d}\sin\frac{p'd}{\hbar} \right|\right\rangle,
\eeq
where the average is over one period of the sinusoidal function.

This last expression can be evaluated analytically for any $k$. However, 
in order to derive a relation between the slit 
separation $d$ and the momentum transfer $D_{M}$ 
we are interested 
only in the minimum over all $k$. 
It is not difficult to verify that the minimum 
occurs for $k=0$, and has the value
\beq
D_{M}^{\rm min}(P_{f},P_{i}) = \frac{d}{\pi\hbar}\int_{0}^{\pi\hbar/d} dp' 
\frac{\hbar}{d}\sin\frac{p'd}{\hbar} = \frac{2\hbar}{\pi d}.
\eeq
Since this is the smallest possible value for $D_{M}$ 
we can thus write a new  momentum-disturbance
 relation
\beq
D_{M}(P_{f},P_{i})  \geq \frac{2\hbar}{\pi d},
\eeq
where the equality can clearly be attained.

\section{\large {\bf Conclusion}}

As explained in the introduction, Heisenberg's measurement--disturbance 
relation (\ref{hur})
between the accuracy of a position measurement $\epsilon_{q}$ 
and the momentum disturbance $\delta p $
applies only when the particle can be treated as being 
initially in a momentum eigenstate.
In any other situation it can be used only heuristically, not 
rigorously. In this paper I have considered one of these other situations, 
a particular one which is of continuing interest, the twin slit. Here the 
particle is prepared in an equal-amplitude superposition of being at both the 
upper and lower slit, separated by a distance $d$. The 
measurement in this case simply distinguishes between these two 
possibilities. In this paper, I have argued that a good measure 
for the momentum disturbance $\Delta p$ is the Monge distance 
$D_{M}(P_{f},P_{i})$ between the 
momentum distributions before and after the measurement. 
Using this measure I have derived  
a new (twin-slit) measurement--disturbance relation:
\beq \label{tsur}
d \times \Delta p \geq \frac{2\hbar}{\pi }.
\eeq

It should be emphasized that $d$ and $\Delta p$  are 
not standard deviations for $q$ and $p$ for any state of the particle. 
Unlike the Heisenberg measurement--disturbance relation (\ref{hur}), 
the relation  (\ref{tsur}) is not derived from \erf{ur}. 
Thus the fact that the right-hand-side of \erf{tsur} is 
different from that of \erf{hur} by a numerical factor of $4/\pi$ is 
not surprising and is of 
no particular significance. However, I 
think it is justifiable to call \erf{tsur} an extension of the Heisenberg 
relation because the origin of the relation is exactly the same, 
namely the conjugate relation between the position and momentum of a 
particle. This conjugate relation is expressed in the Fourier 
transform which takes one from the 
position representation to the momentum representation, and which 
gives rise to the oscillations of period $2\pi\hbar/d$ 
in the initial momentum distribution
of \erf{imd} which are erased by the measurement. The smaller the 
separation between the two slits, the more accurate the position 
measurement must be to resolve them, and the larger the momentum 
transfer. 

With regard to the use of the Monge distance as a measure of the 
momentum transfer, it might be questioned whether this has any ``physical'' 
(rather than mathematical) justification 
for this measure. It turns out that the answer is yes, if one is 
prepared to accept the Bohmian  
interpretation \cite{Boh52} of quantum mechanics as 
physical. This idea is explored extensively in Ref.~\cite{WisBohm}, 
where I show that the individual trajectories taken by particles 
under Bohmian mechanics can be traced both with and without a 
measurement for the twin-slit case. In the far-field, the velocities of 
these particles can be compared in these two cases, and the momentum 
change caused by the position measurement computed. If one defines 
$\Delta p$ to be the absolute value of the momentum change, 
averaged over all of the possible initial starting points of the 
particle, then one finds that this measure obeys exactly the 
inequality (\ref{tsur}). 

Having mentioned the Bohmian interpretation, 
it is worth re-emphasizing that the results of this paper  
are completely
independent of one's interpretation of quantum mechanics. 
The momentum disturbance $\Delta p$ is just the 
integral of the absolute value of the difference between the 
fiducial momentum distributions with and without 
the measurement. This integral is no harder to calculate from 
experimental data than the 
standard deviation of the momentum distribution which appears in the 
original Heisenberg measurement--disturbance relation (\ref{hur}).

Finally, one might wonder whether the result in this paper points the 
way towards a more general measurement--disturbance relation which 
would hold not only for the momentum eigenstate and twin-slit cases but 
for all initial conditions. Unfortunately I think the answer is no. 
The problem is not one of the definition of momentum transfer (the one 
used in this work would seem to be generally applicable), but of the 
measurement error. As 
a trivial example, if the initial state is sufficiently well 
localized in position then a position measurement of finite error 
may have no effect on the state whatsoever, so that 
there will be no momentum disturbance $\Delta p$. What distinguishes 
the momentum eigenstate and twin-slit cases is that 
in these cases a position 
measurement does have a clear effect on the particle. In the first 
case the measurement error $\epsilon_{q}$ 
can be chosen arbitrarily; in the second 
case it is the slit separation $d$ which is the relevant length scale 
(providing the two slits are distinguished by the measurement). 
It may be possible to work out measurement--disturbance relations for 
other particular examples, but it seems doubtful that they would 
supply any more insight 
than can be obtained from considering the two obvious cases.

\acknowledgments

I am indebted to D.F. Walls for expressing (a number of years ago) 
the view that there should 
be a way to work out the momentum transfer just by examining the 
the momentum distributions before and after the measurement. I would 
also like to thank D.A. Rice for bringing to my attention 
the quotation I use from Heisenberg's 1930 book, and 
M.J. Gagen for comments on the first draft of this manuscript.
This work was supported by the 
Australian Research Council.

}

\end{document}